\documentstyle[11pt,aaspp4,flushrt]{article} 

\begin{document}

\def\spose#1{\hbox to 0pt{#1\hss}}
\def\lta{\mathrel{\spose{\lower 3pt\hbox{$\mathchar"218$}}
     \raise 2.0pt\hbox{$\mathchar"13C$}}}
\def\gta{\mathrel{\spose{\lower 3pt\hbox{$\mathchar"218$}}
     \raise 2.0pt\hbox{$\mathchar"13E$}}}

\def\msun{{\rm M}_\odot}
\def\rsun{{\rm R}_\odot}
\def\lsun{{\rm L}_\odot}
\def\teff{T_{\rm eff}}
\def\dd{{\rm d}}
\def\zr{\zeta_{\rm R}}
\def\RL{R_{\rm L}}
\def\m2i{M_{2 {\rm i}}}

\def\zs{\zeta_{s}}
\def\half{{1\over2}}
\def\dJJ{{\dot J\over J}}
\def\dMM{{\dot M_2\over M_2}}
\def\tKH{t_{\rm KH}}
\def\eck#1{\left\lbrack #1 \right\rbrack}
\def\rund#1{\left( #1 \right)}
\def\wave#1{\left\lbrace #1 \right\rbrace}

\slugcomment{\bf Accepted for publication in ApJ Letters (May 1997).}

\title{THE TRANSIENT NATURE OF GRO~J1655-40 \\ 
AND ITS EVOLUTIONARY STATE} 

\author{U. Kolb \altaffilmark{1}, A. R. King \altaffilmark{1}, 
H. Ritter \altaffilmark{2}, and J. Frank
\altaffilmark{3, 4}
}

\altaffiltext{1} {Astronomy Group, University of Leicester, 
Leicester LE1 7RH, U.K.; uck@star.le.ac.uk; ark@star.le.ac.uk}

\altaffiltext{2} {Max--Planck--Institut f\"ur Astrophysik, 
Karl--Schwarzschild--Str.~1, D~85740 Garching, Germany; \\
hsr@MPA-Garching.MPG.DE}

\altaffiltext{3} {Space Telescope Science Institute, 3700 San Martin
Dr., Baltimore, MD 21218, USA}

\altaffiltext{4} {Department of Physics and Astronomy, Louisiana State
University,   Baton Rouge, LA 70803-4001, USA;
frank@rouge.phys.lsu.edu}

\begin{abstract}
We consider the evolutionary state of the black--hole X--ray source
GRO~J1655--40 in the context of its transient nature. 
Recent optical observations show that the donor in GRO~J1655--40
is an intermediate--mass star ($\simeq 2.3 \msun$) crossing the 
Hertzsprung gap. Usually in such systems the donor's radius expansion
drives a near-Eddington or super--Eddington mass transfer
rate which would sustain a persistently bright accretion disk. 
We show that GRO~J1655--40 is close to a narrow parameter range 
where disk instabilities can occur. This
range corresponds to a short--lived evolutionary stage where the
secondary's radius expansion stalls (or reverses), with a
correspondingly lower mass transfer rate.  
If GRO~J1655--40 belongs to this class of transients
the predicted accretion rates imply large populations of luminous
persistent and transient sources, 
which are not seen in X--rays. 
The transient nature of the related system GRS~1915+105 may 
reflect spectral variations in a bolometrically persistent source rather 
than a genuine luminosity increase.
\end{abstract}


\newpage

\section{INTRODUCTION}

In a recent series of papers we have considered the evolution of
low--mass X--ray binaries (LMXBs), and in particular
the question of which LMXBs should be transient
according to the disk instability model. 
To have low enough mass transfer rates for transient behavior, 
short--period  ($P \lta 1$~d) neutron star systems require 
special formation conditions (King, Kolb \& Burderi 1996, hereafter
KKB). These formation conditions can indeed be shown to hold in a
significant fraction of neutron--star systems (King \& Kolb 1997).
By contrast almost all short--period black--hole systems will be
transient (King, Kolb \& Szuszkiewicz 1997, hereafter KKS).
Longer--period LMXBs whose evolution is driven by the nuclear
expansion of an evolved low--mass companion with a degenerate helium
core are almost always transient, regardless of whether they contain a
black hole or a neutron star (KKS and King et al.\ 1997). 

The recently discovered soft X--ray transient
GRO~J1655--40 (Orosz \& Bailyn 1997) belongs to none of these
classes. The companion orbits a  $7\msun$ black hole with a period of
2.62 days, and
has a rather large mass ($\simeq 2.3 \msun$), too high for the simple
core--envelope picture used by KKS, King et al.\ (1997) to hold.  
Its spectral type (F3--F6; effective temperature $T_{\rm
eff} \simeq 6500$~K) actually implies a star in the process of
crossing the Hertzsprung gap. 
Since here the stellar radius $R$ expands significantly on
the star's thermal time $\sim 10^7$~yr, one would expect mass transfer rates
$\dot M \sim 10^{-7}$~$\msun$~yr$^{-1}$, which are too high for 
disk instabilities to occur, particularly when one allows for
irradiation of the disk by the central accreting source 
(van Paradijs 1996). 

Nevertheless GRO~J1655--40 clearly shows transient behavior, though in
a rather complex form: the BATSE detectors aboard the Compton GRO observed 
distinct hard X--ray outbursts in July 1994 (when the source was
discovered), as well as September 1994, November 1994, March 1995 and July 1995
(Harmon et al.\ 1995, Tavani et al.\ 1996). The 1994 outbursts were
associated with radio flaring and apparent
superluminal motion of radio plasmoids (Hjellming \& Rupen 1995),
giving the system distance as $d=3.2$~kpc. After a
phase of true X--ray quiescence a soft X--ray outburst was
observed by RXTE in April 1996 (Levine et al.\ 1996), again
accompanied by brightening in other wavebands.

Accordingly we consider the evolution of GRO~J1655--40 in detail
here. We apply results of computations of LMXBs with donors crossing the
Hertzsprung gap, details of which will be presented 
elsewhere (Kolb 1997). We shall see that a short transient phase is
indeed possible, although the observed parameters of GRO~J1655--40 place
it slightly outside our currently computed instability
strip. Uncertainties in the input stellar physics 
and perhaps in the observationally determined value for the effective
temperature probably allow the
source to lie within the strip.  
Since the instability strip is narrow, we consider the consequences of the
high implied space density of systems like GRO~J1655--40.

\section{TRANSIENT SYSTEMS IN THE HERTZSPRUNG GAP}

A star crossing the Hertzsprung gap has finished core hydrogen
burning, but not yet ignited core helium burning. 
Mass transfer from such a donor
is known as Case B. This case is well studied (e.g. Kippenhahn et al.\
1967, Pylyser \& Savonije 1988, Kolb \& Ritter 1990, De~Greve 1993),
particularly in the context of Algol evolution. The main difference
for black hole LMXBs is that the donor is the less massive
star. Elsewhere we shall present extensive computations of this case
(Kolb 1997), in 
which  we distinguish transient from persistent LMXBs by comparing the
mass transfer rate $\dot M$ with the critical rate $\dot M_{\rm crit}$
at which hydrogen is just ionized at the outer disk edge. A system is 
transient if $\dot M < \dot M_{\rm crit}$. 

We can summarize these results as follows.
 
Figure~1 gives the evolutionary tracks of the donor stars of selected
sequences on an HR diagram with parameters chosen to lie close to
GRO~J1655--40. The positions of single stars of fixed mass are also
shown. At any time the luminosity and effective temperature of the
donor is roughly the same as that of a single star of the same mass   
and radius. 
This explains why Orosz \& Bailyn (1997) found parameters for the
secondary in GRO~J1655--40 which are `consistent with a $2.3\msun$
single star evolution'. However, the donor stars are not in  
thermal equilibrium, and their evolution through any point on the
figure depends on their previous history. 
In particular for much of the time the mass transfer rates for the
binaries are of order $M_2/t_{\rm th}$, where $t_{\rm th}$ is the
thermal timescale of the donor (of order the Kelvin--Helmholtz time)
at the start of mass transfer, and thus 
depend strongly on the mass $\m2i$ of the donor at that initial
time. The portions of the tracks in thick linestyle show
where the systems are transient according to the criterion of KKS. As
can be seen, in addition to the expected transient phase when the
donors are close to the low--mass giant branch, there is a short
transient phase when the star is still in the Hertzsprung gap. 
This phase occurs because the stellar radius expands very little (or
even contracts) at this point; this in turn is a result of the complex
structural change brought about by the switch from thick to thin shell
H--burning. 
The radius slowdown always occurs at the same effective temperature
as for a single star with the donor's initial mass, but at roughly the
same radius as for a single star with the donor's actual mass.

These findings can be conveniently generalized to arbitrary 
$\m2i$ in a plot of orbital period versus donor mass
(Fig.~2). Conservative evolution with constant binary mass $M$ is towards
longer $P$ and smaller   
$M_2$ along the curves $P M_2^3 (M-M_2)^3=$~const, while along an
evolution with mass loss $|\dd P/\dd M_2|$ is larger 
(see the tracks shown in the Figure). 
The hatching on the Figure relates to the
transient/persistent nature of the systems:  
in the unhatched region all black--hole systems are persistent, and in
the narrow hatched strip bisecting this zone all the systems are
transient. This is the phase where the radius expansion slows. Because
this only depends on the current stellar mass, it is insensitive to
the initial mass $\m2i$, 
unlike most other properties of the evolutionary sequences. 
Most systems in the large hatched region are also transient (this is
the phase where the donor becomes a low--mass giant), although systems
evolving into this region from the unhatched `transient exclusion
zone' (thus with large initial donor mass) remain persistent until
$M_2$ has dropped somewhat.
The transient strip terminates at $M_2 \simeq 3.5\msun$, as for more
massive donors $\dot M > \dot M_{\rm crit}$ even when the
radius expansion slows.

\section{APPLICATION TO GRO~J1655--40}

Taken at face value, the system parameters of GRO~J1655--40 
as derived by Orosz \& Bailyn (1997) place it in a region of the $P -
M_2$ plane where it should be persistent rather than transient, with a
mass transfer rate close to the Eddington limit (if the initial donor
mass was $\la 2.5\msun$) or super--Eddington (if $\m2i \ga 2.5\msun$).
In this case the observed outbursts
and variability must presumably be caused by instabilities in the accretion
flow at Eddington- or super--Eddington mass transfer rates. 
Small variations in the accretion rate could have a significant effect
on the appearance of the source by changing its effective photosphere.  
However, GRO~J1655--40 is classified as a transient, and indeed,
the X--ray outbursts show characteristic features consistent 
with the operation of an underlying disk instability.
Simple scalings for the outburst duty cycle and
recurrence time of soft X--ray transients, taking into account the
effect of irradiation on the outburst (King \& Ritter 1997), suggest
an irregular outburst pattern in long--period systems like
GRO~J1655--40: a series of outbursts with short recurrence time (the
presently observed behavior) may alternate with a much longer
quiescence (the behavior before the first detection in 1994). The
delay between the optical and X--ray outburst observed in April 
1996 (Orosz et al.\ 1997) is in close analogy to the UV delay observed
in dwarf novae. In terms of the disk instability model such 
delays occur naturally whenever the disk does not extend all the way
to the white dwarf surface, or the innermost stable orbit in the case
of a black hole accretor (e.g. Livio \& Pringle 1992, King 1997). Hameury et
al.\ (1997) suggest the presence of a ``hole'' in the disk of
GRO~J1655--40 by postulating that the disk flow gives way to advection
at some inner radius.

In the light of this evidence it is intriguing that the observed
system parameters put GRO~J1655--40 quite close to the
short transient phase that similar systems encounter when the secondary
evolves through the Hertzsprung gap. If we assume that GRO~J1655--40
in reality actually falls into the narrow transient strip,
its outburst behavior is consistent with the disk instability
model for soft X--ray transients. This would require some error in the
position of the transient strip or of the system. The precise location of the
transient strip in the HR diagram is determined by the effective
temperature where single stars in the mass range $2-4\msun$
temporarily slow or reverse their expansion from the main sequence to
the first giant branch. This effective temperature (and hence the
radius and luminosity) depends on details of the input physics of the
evolution code, particularly opacities and the treatment of
convection. 
In addition, it is conceivable that the observational determination of
$\teff$ might yet change. A value $\teff \la 5600$~K would place
GRO~J1655--40 in the predicted transient strip. In Orosz \& Bailyn's
spectral classification procedure 
the corresponding spectral type G2IV achieves a minimum r.m.s.\ value 
which is only $\sim 10\%$ larger than that of the best--fit type F4IV. 
Roche geometry implies with $\teff = 5600$~K a secondary luminosity
$\simeq 20\lsun$. This would require a color excess $E(B-V) \la 1.0$
(rather than 1.3) for consistency with the observed mean V magnitude
(17.12 mag) and distance (3.2~kpc).
A further uncertainty may arise from the fact that the accretion disk still
contributes $\la 10\%$ of the V flux during the extended X--ray
quiescence in 1996, and as much as $50\%$ in 1995 (Orosz \& Bailyn 1997).

Although this transient phase is very short ($\simeq 1$~Myr if
$\m2i=2.5\msun$), it is quite plausible
that such a system would nevertheless be observable, because its
outbursts would be extremely luminous. 
The actual transfer rate in the transient phase depends on the initial
orbital separation. The closer the initial
separation was to the present value, the lower the transfer rate. 
With a typical transient duty cycle  
$\lta 10^{-2}$ the instantaneous accretion rate can be as high as
$\dot M \sim 10^{-8} - 10^{-6}$~$\msun$~yr$^{-1}$, implying
luminosities up to $10^{38} - 10^{40}$~erg~s$^{-1}$. 
This is consistent with the outburst X--ray luminosity $\ga
10^{38}$~erg~s$^{-1}$ inferred for GRO~J1655--40 from observations
(e.g.\ Tanaka \& Shibazaki 1996).
X--ray luminosities of anything approaching this order 
are visible throughout the Galaxy, as observations of high--mass
X--ray binaries show. 
However, just as for those systems, the price one
pays for asserting the visibility of a short--lived evolutionary phase
is the requirement of a very high birth rate. Here we would need 
a rate of order 1 per $10^4$ yr. 
The giant branch transient phase is $10-20$ times longer than the
Hertzsprung gap transient phase. The detection probability of the 
correspondingly higher number of giant branch transients with
intermediate--mass donors may nevertheless be comparable to that of
the Hertzsprung gap transients as the duty cycle decreases with
increasing orbital period (King \& Ritter 1997). 
A more serious problem is that the required high formation rate
implies about $500$ bright persistent accreting sources in the
Hertzsprung gap, in apparent conflict with observation. In the next    
section we examine possible resolutions of this disagreement.

\section{DISCUSSION}

The disk instability picture provides a very clean separation of LMXBs
into transient and persistent sources, provided that one takes due account
of the effects of irradiation by the central accretor (van Paradijs 1996).
This is strong circumstantial evidence in its favor. 
Accordingly one would be very reluctant to abandon it simply because 
the application of this picture to the transient source
GRO~J1655--40 leads to an apparent overpopulation of bright accreting
sources in the Galaxy. 

The resolution of this conflict is probably related to the fact 
that most systems populating the $P-M_2$ plane in Fig.~2
are extremely luminous. The persistent sources with $\m2i \la 2.5\msun$
would be close to the Eddington limit for a $7\msun$ black hole (and
super--Eddington if $\m2i\ga 2.5\msun$), while 
with typical duty cycles, the outbursts of the transient systems 
could be still brighter. 
This raises the possibility of significant mass  
loss, in two forms. GRO~J1655--40 and another system, GRS~1915+105,
are seen to lose 
mass in relativistic jets (Hjellming \& Rupen 1995, Mirabel \&
Rodr\'iguez 1994), although there is no consensus on the actual amount
of mass lost.
Further, it is theoretically possible (e.g. Begelman, McKee \& Shields
1983, Czerny \& King 1989) for most of the transferred mass to
be lost from the outer parts of the accretion disk, driven by the
luminosity of the accreting fraction. There is indirect evidence that
much of the mass transferred in neutron--star LMXBs is lost, since the
remnant pulsars always have rather modest masses (see King \& Kolb
1997 for a discussion). 

This mass loss may prevent the bright systems appearing as X--ray
sources, either because much of the mass never accretes to the compact
primary, or because the mass loss itself shrouds the system and
degrades the X--rays.  
Alternatively, the accretion rates may be too high for X--ray 
production. The effective photosphere for the accretion flow may
actually radiate at much longer wavelengths for most of 
the time.  
Since most of these sources are
probably rather distant and heavily reddened, the infrared is the most
likely place to detect them.  
If GRO~J1655--40 is indeed a genuine soft X--ray transient in the
Hertzsprung gap, then either of these possibilities would reconcile
theory and observation.

In this connection it is interesting that GRS~1915+105,
also classified as a soft X--ray transient,
has been persistently bright in X--rays since 1992. At
its minimum X--ray luminosity $\sim 5\times 10^{38}$~erg~s$^{-1}$
(e.g. Belloni et al.\ 1997) this is far longer than can be explained
by a disk instability. It may be that this source, too, is actually one of the
`persistent' sources predicted by Fig.~2. Certainly its current mass 
accretion rate $\sim 10^{-7}$~$\msun$~yr$^{-1}$ is consistent with that
expected, although the heavy interstellar reddening
makes the nature of the companion very uncertain (Mirabel et al.\ 1997).
The current (post--1992) state may correspond to a 
{\it lower} accretion rate, so that the X--rays become temporarily 
visible (see also Belloni et al.\ 1997). Infrared monitoring of this
source should provide more information about this possibility, in that
the X--rays and infrared should be anticorrelated.

This
work was partially supported by the U.K.\ Particle Physics and Astronomy
Research Council (PPARC) and by grant NAG 5-3082 to LSU.
ARK acknowledges support as a PPARC Senior Fellow. JF thanks the Space
Telescope Science Institute for its hospitality. 
We thank the anonymous referee for useful comments.

{}

\clearpage


\bigskip



\begin{figure}
\plotone{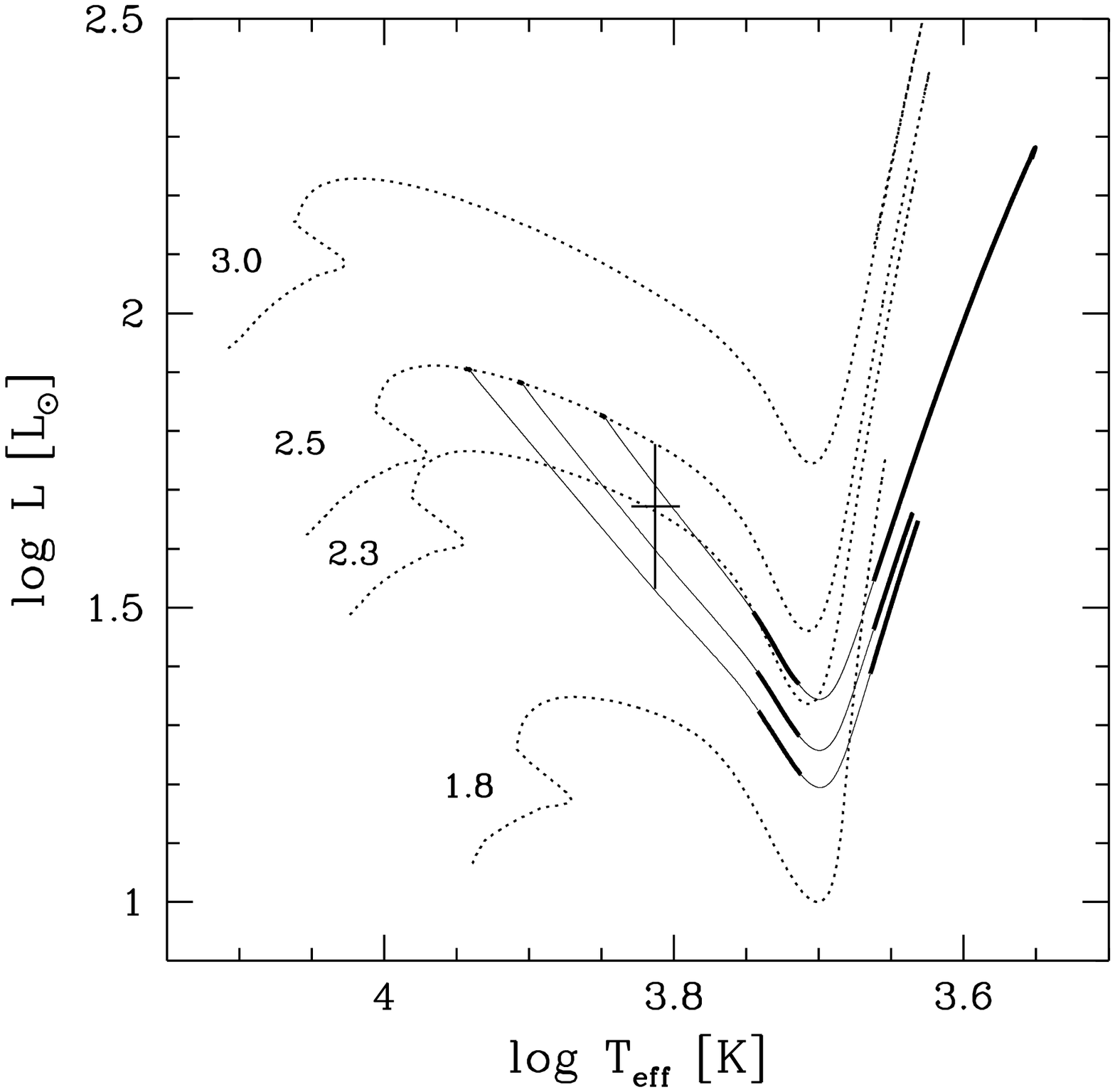}
\caption{
HR diagram with various evolutionary sequences. Dotted: single stars
(labelled with stellar mass in $\msun$). Solid: donor stars in 
black hole LMXBs (initial black hole mass $6.8\msun$, initial donor mass
$2.5\msun$, initial separation $a=13.2$, $15.2$ and 
$18.5\rsun$). Conservative mass transfer was assumed.
Different linestyle indicates where the binary appears as 
a transient (solid) and persistent source (thin). 
The $2.5\msun$ single star evolution and the $a=18.5\rsun$ sequence
are shown up to ignition of core He burning, all other sequences have
been terminated earlier. For further details see Kolb (1997). The
cross marks the observed parameters of the donor star in GRO~J1655--40
(Orosz \& Bailyn 1997).  
}
\end{figure}

\begin{figure}
\plotone{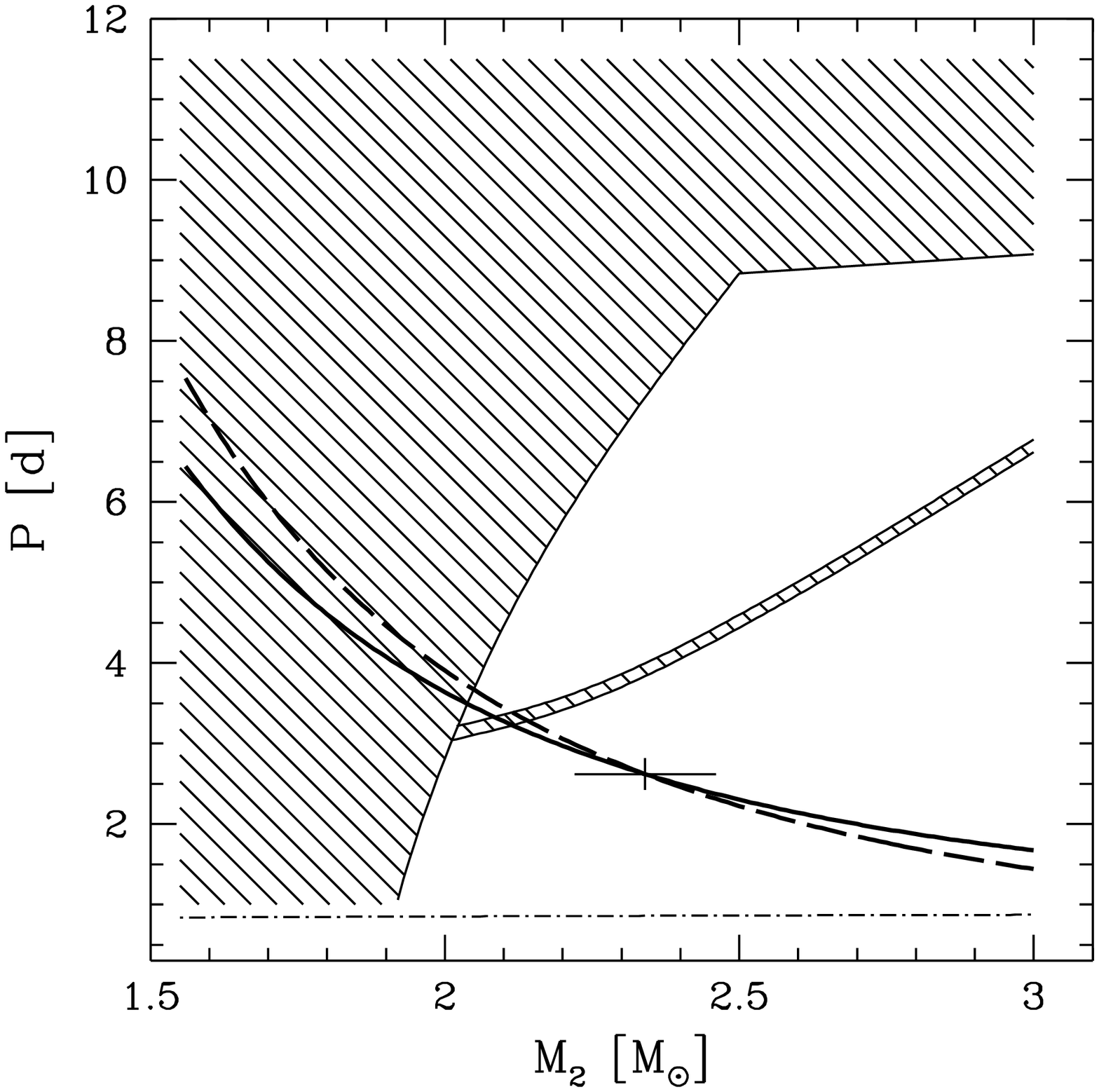}
\caption{
Orbital period -- secondary mass ($P-M_2$) plane for black hole LMXBs, with 
the exclusion zone for transients (unhatched). A black hole mass of
$8\msun$ was assumed.  
Systems in the unhatched region are always persistent; systems in the
hatched narrow instability strip are always transient. Systems 
born in the large hatched area are always transient, but systems
evolving from the unhatched into the large hatched area 
remain persistent for some time (depending on the evolutionary
prehistory) before they, too, become transient.
The dash--dotted line (at the bottom of the figure)
is the binary period when the secondary
fills its Roche lobe on the 
terminal age main--sequence. 
The cross indicates the observed
system parameters of GRO J1655--40 (Orosz \& Bailyn 1997). The solid
curves represent evolutionary tracks assuming
conservative evolution (full line) and evolution with constant black
hole mass (mass leaving the system carries the black hole's
specific orbital angular momentum; dashed). 
}
\end{figure}

%
%


\clearpage



\begin{thebibliography}{}

\bibitem{} Begelman M.C., McKee C.F., \& Shields G.A. 1983, ApJ, 271, 70

\bibitem{} Belloni T., M\'endez M., King A.R., van der Klis M., \&
van Paradijs J. 1997, ApJ, 479, L145 

\bibitem{} Czerny M., \& King A.R. 1989, MNRAS, 241, 839

\bibitem{} 
De~Greve J.P. 1993, A\&AS, 97, 527

\bibitem{}
Hameury J.-M., Lasota J.-P., McClintock J.E., \& Narayan R. 1997, ApJ,
submitted

\bibitem{}
Harmon B.A., Wilson C.A., Zhang S.N., Paciesas W.S., Fishman G.J.,
Hjellming R.M., Rupen M.P., Scott D.M., Briggs M.S., \& Rubin
B.C. 1995, Nature, 374, 703

\bibitem{}
Hjellming R.M., \& Rupen M.P. 1995, Nature, 375, 464

\bibitem{}
King A.R. 1997, MNRAS, in press

\bibitem{}
King A.R., \& Kolb U. 1997, ApJ, 481, 918

\bibitem{}
King A.R., \& Ritter H. 1997, MNRAS, submitted

\bibitem{}
King A.R., Kolb U., \& Burderi L. 1996, ApJ 464, L127 (KKB)

\bibitem{}
King A.R., Kolb U., \& Szuszkiewicz E. 1997, ApJ, 488, Oct.~10 issue (KKS)

\bibitem{}
King A.R., Frank J., Kolb U., \& Ritter H. 1997, ApJ, in press

\bibitem{}
Kippenhahn R., \& Weigert A. 1967, Zeitschrift f\"ur
Astrophysik, 65, 251

\bibitem{}
Kolb U. 1997, MNRAS, submitted

\bibitem{}
Kolb U., \& Ritter H. 1990, A\&A, 236, 385

\bibitem{}
Levine A.M., Bradt H., Cui W., Jernigan J.G.,
Morgan E.H., Remillard R., Shirey R.E., \& Smith D.A. 1996, ApJ, 469,
L33 

\bibitem{}
Livio M. \& Pringle J.E. 1992, MNRAS, 259, 23P

\bibitem{}
Mirabel I.F., \& Rodr\'iguez L.F. 1994, Nature, 371, 46

\bibitem{}
Mirabel I.F., Bandyopadhyay R., Charles P.A., Shabaz T., \&
Rodr\'iguez L.F. 1997, ApJ, 477, L45 

\bibitem{}
Orosz J.A., \& Bailyn C.D. 1997, ApJ, 477, 876 

\bibitem{}
Orosz J.A., Remillard R.A., Bailyn C.D., \& McClintock J.E. 1997, 
ApJ, 478, L83

\bibitem{}
Pylyser E.H.P., \& Savonije G.J. 1988, A\&A, 191, 57

\bibitem{}
Tanaka Y., \& Shibazaki N. 1996, ARA\&A, 34, 607

\bibitem{}
Tavani M., Fruchter A., Zhang S.N., Harmon B.A., Hjellming R.N., Rupen
M.P., Bailyn C., \& Livio M. 1996, ApJ, 473, L103

\bibitem{}
Van Paradijs J. 1996, ApJ, 464, L139

\end{thebibliography}
\end{document}